\documentstyle[12pt]{article}

\def\endpage{\vfill\eject}

\newcommand{\AmS}{{\protect\the\textfont2
\renewcommand{\thesection}{\Roman{section}}
  A\kern-.1667em\lower.5ex\hbox{M}\kern-.125emS}}
\begin{document}
\rightline {\bf DFTUZ/95/12}
\rightline {to appear in Phys. Lett. B}
\vskip 2. truecm
\centerline{\bf CHIRAL SUSCEPTIBILITIES IN NONCOMPACT QED: A NEW}
\centerline{\bf DETERMINATION OF THE $\gamma$ EXPONENT AND THE 
CRITICAL COUPLINGS}
\vskip 2 truecm
\centerline { V.~Azcoiti$^a$, G. Di Carlo$^b$, A. Galante$^{c,b}$, 
A.F. Grillo$^d$, V. Laliena$^a$ and C.E. Piedrafita$^b$}
\vskip 1 truecm
\centerline {\it $^a$ Departamento de F\'\i sica Te\'orica, Facultad 
de Ciencias, Universidad de Zaragoza,}
\centerline {\it 50009 Zaragoza (Spain).}
\vskip 0.15 truecm
\centerline {\it $^b$ Istituto Nazionale di Fisica Nucleare, 
Laboratori Nazionali di Frascati,}
\centerline {\it P.O.B. 13 - Frascati 00044 (Italy). }
\vskip 0.15 truecm
\centerline {\it $^c$ Dipartimento di Fisica dell'Universit\'a 
dell'Aquila, L'Aquila 67100 (Italy)}
\vskip 0.15 truecm
\centerline {\it $^d$ Istituto Nazionale di Fisica Nucleare, 
Laboratori Nazionali del Gran Sasso,}
\centerline {\it Assergi (L'Aquila) 67010 (Italy). }
\vskip 3 truecm
\centerline {ABSTRACT}
We report the results of a measurement of susceptibilities in noncompact 
$QED_4$ in $8^4, 10^4$ and $12^4$ lattices. Due to the potentialities of the 
$MFA$ approach, we have done simulations in the chiral limit which are 
therefore free from arbitrary mass extrapolations. Our results in the Coulomb 
phase show unambiguously that the susceptibility critical exponent $\gamma=1$ 
independently of the flavour symmetry group. The critical couplings extracted 
from these calculations are in perfect agreement with previous determinations 
based on the fermion effective action and plaquette energy, and outside the 
predictions of a logarithmically improved scalar mean field theory by eight 
standard deviations.

\vfill\eject

\par

Non compact Electrodynamics in four dimensions $(QED_4)$ has a strongly 
coupled phase where chiral symmetry is spontaneously broken. The strong 
coupling phase is connected to the weak coupling phase through a continuous 
phase transition characterized by a critical point where a continuum 
limit can be defined. The possible existence of a non trivial continuum 
limit for this model is an appealing question which has attracted much 
attention in the last years and whose answer is not well known yet.
\par
The main difficulty in establishing the nature of the continuum limit of this 
model comes from the fact that it can not be investigated in perturbation 
theory. The possible existence of a non gaussian fixed point implies that 
operators of dimension higher than four could become nonperturbatively 
renormalizable \cite{BARDEEN} and therefore dimensional analysis does not 
help us to establish what the relevant couplings are.
\par
Due to the high complexity of this problem, the first approaches  
have been concentrated in the determination of thermodynamical 
quantities, critical exponents and their comparison with 
the predictions of a mean field theory 
plus logarithmic corrections. Several numerical methods have been used 
by different groups to simulate dynamical fermions. In refs. 
\cite{HOMBRES}, \cite{TED} 
the hybrid Monte Carlo algorithm was used whereas in our previous work 
\cite{NOS} the $MFA$ approach \cite{MFA} was employed. 
Even if the numerical results 
obtained by all these groups are in reasonable agreement, their physical 
interpretations are in conflict. 

In \cite{TED} the value of the critical coupling is left as a free parameter 
in a fit of the numerical results with a logarithmically improved scalar 
mean field 
theory. The five parameters fit accommodates quite well the experimental 
data and the prediction of the fit for the critical coupling $\beta_c$ in 
the four flavour theory is $\beta_c=0.186(1)$. These results have been 
interpreted by the authors of \cite{TED} as strong evidence of a trivial 
continuum limit for this model. Notice however that if as recently 
argued \cite{SACHA}, triviality manifests in a different way in scalar 
and fermionic models, the logarithmically improved mean field test of 
\cite{TED} would not be justified.

Conversely, the results reported in \cite{HOMBRES} 
and \cite{NOS} for the critical couplings, both of them in very good 
agreement, show a value for $\beta_c$ 
larger than $0.186$. The excellent agreement 
between the results of \cite{HOMBRES} and \cite{NOS} 
should be of great physical 
relevance since both simulations are perfectly uncorrelated. In fact 
these two groups use different approaches to simulate dynamical 
fermions, different lattice sizes, different operators and different 
bare fermion masses (the simulations in \cite{NOS} were done in the chiral 
limit).

Since the numerical determination of critical exponents is very sensitive 
to the value of the critical coupling $\beta_c$ and that values of 
$\beta_c>0.186$ could produce critical exponents outside of the 
mean field range, it is very important to have more independent 
measurements of $\beta_c$. 

We report here the results of a measurement of the 
longitudinal, transverse and non linear susceptibilities in $QED_4$ 
with $0, 2$ and $4$ flavours. This is the first time that susceptibilities 
in noncompact $QED$ are calculated in the chiral limit (Coulomb phase) and 
the results are certainly encouraging. In fact 
the values obtained for the critical couplings 
are in perfect agreement with previous determinations 
based on measurements of the fermionic effective action and plaquette 
energy \cite{NOS} and therefore we confirm again from these results that 
$\beta_c$ in the four flavour model is definitely larger than $0.186$. These 
measurements show also unambiguously that the susceptibility critical 
exponent $\gamma$ is 1, independently of the flavour symmetry group. 
Furthermore 
the value of $\beta_c$ in the quenched model extracted from the susceptibility 
is also in perfect agreement with previous determinations of it from the 
computation of the chiral condensate \cite{MARIA} and larger than the value 
$\beta_c=0.244$ in the zero flavour model extracted from 
the singular behavior of the fermionic action \cite{NOS}, 
the last being also in perfect agreement with the critical $\beta$ obtained 
from the monopole percolation transition \cite{HOMBRES}.

We have done calculations in $8^4, 10^4$ and $12^4$ lattices and dynamical 
fermions are simulated by mean of the $MFA$ method, the only available 
numerical approach which allows to perform realistic numerical 
simulations in the chiral limit. $MFA$ (Microcanonical Fermionic Average) 
has been extensively discussed \cite{MFA}  
and tested \cite{SCH} in the literature. Let us however remember what are 
the main ingredients of the $MFA$ approach.

The basic idea in $MFA$ is the exploitation of the physical
equivalence between the canonical and the microcanonical
formalism, in our case the introduction of an explicit dependence
on the energy in the computation of the partition function.
Indeed it can be written as follows:

$$
Z(\beta,m)=\int dE n(E) e^{6V\beta E} \overline{\det} \Delta(E,m)
\eqno(1)$$

\noindent
where 

$$n(E)=\int [d A_\mu] \delta(6VE-S_G[A_\mu])
\eqno(2)$$

\noindent
is the density of states at fixed energy $E$ and

$$
\overline{\det} \Delta(E,m)={\int [d A_\mu] \delta(6VE-S_G[A_\mu])\det 
\Delta[A_\mu,m] \over n(E) } 
\eqno(3)$$

\noindent
is the fermionic determinant averaged over field configurations of 
fixed energy $E$. Note that $\overline{\det} \Delta(E,m)$ 
does not depend on $\beta$.
The fermionic determinant for a single configuration
is a polynomial in the mass:
$$
\det \Delta[A_\mu,m] =\sum_n  C_n[A_\mu] m^n
\eqno(4)$$

A modified Lanczos algorithm \cite{LANCZOS} is used in order
to obtain the complete set of eigenvalues of the
massless fermion matrix, and then we can reconstruct the 
fermionic determinant at any value of the fermion mass $m$.

The most general expression for the vacuum expectation value of any 
operator $O$, after integration of the Grassmann variables, can be 
written as

$$<O> = {\int dE n(E){\overline{O\det}\Delta\over{\overline{\det}\Delta}}
 e^{-6V\beta E-S^{F}_{eff}(E,m)} 
\over {\int dE n(E) e^{-6V\beta E-S^{F}_{eff}(E,m)}}} 
\eqno(5)$$

\noindent
where $S^{F}_{eff}(E,m)$ in (5) is the fermion effective action defined 
as 

$$S^{F}_{eff}(E,m) = -\log \overline{\det} \Delta(E,m)
\eqno(6)$$

\noindent
and $\overline{O\det}\Delta$ means the mean value of the product of 
the operator $O$ times the fermionic determinant, computed over gauge 
field configurations at fixed energy $E$.

Since we are interested here in the computation of susceptibilities in 
the chiral limit, we 
will use the previous expression for the particular cases in which $O$ 
is the longitudinal, transverse and non linear susceptibility.

In the Coulomb phase, characterized by a ground state invariant under 
chiral transformations in the chiral limit, the longitudinal and 
transverse susceptibilities are equal except a sign. They can be 
computed by taking for the operator $O$ the expression

$$O={2\over V} \sum_i  {1\over{\lambda^2_i}}
\eqno(7)$$

\noindent
where the sum in (7) runs over all positive eigenvalues of the massless 
Dirac operator.

The transverse susceptibility $\chi_T$ in the broken phase diverges always 
because of the Goldstone boson. The longitudinal susceptibility $\chi_L$ 
on the other hand, can not be computed in the broken phase using equation 
(7) since in this phase the ground state is not invariant under chiral 
transformations. In fact if we work directly in the chiral limit, we 
average over all ground states. In other words there are extra contributions 
to $\chi_L$ in the massive case, which vanish in the chiral limit in the 
Coulomb phase because of the realization of chiral symmetry, but giving 
an important contribution to $\chi_L$ in the broken phase. The standard 
procedure in the broken phase is then to compute the susceptibility at 
non zero fermion mass and to extract its value in the chiral limit by 
fermion mass extrapolations.

Since the susceptibilities in the symmetric phase are free from mass 
extrapolations, we will devote the largest part of this paper to the 
analysis of our results in the Coulomb phase which are the more relevant 
from a physical point of view.

The results in the $8^4$ lattice are the best from a statistical point of 
view since in this lattice we have diagonalized $3000$ completely 
uncorrelated gauge configurations for each value of the energy and 
repeated this process for $17$ different energies between $0.5$ 
and $1.2$. In the $10^4$ lattice, $300-400$ gauge configurations were 
diagonalized at each energy, whereas our results in the $12^4$ lattice are 
statistically the poorest. 

In Fig. 1 we have plotted the inverse susceptibility 
in the chiral limit against $\beta$ for the four-flavour model in a $10^4$ 
lattice. The continuous 
line is a fit of all the points for $\beta>0.210$ with a function

$${\chi}^{-1} = {c\over{(\beta_c+r)}} {(\beta_c - \beta)\over{(\beta+r)}}
\eqno(8)$$

The use of this fitting function is strongly suggested by the results of 
the inverse susceptibility as a function of the energy (Fig. 2), which 
are extremely well fitted in the Coulomb phase by a straight line,  
and by the relation between the mean plaquette energy and $\beta$ 
in the massless case \cite{NOS} which is extremely well reproduced by the  
function

$$E = {1\over{4(\beta+r)}}
\eqno(9)$$

\noindent
the value of $r$ depending on the flavour number.

Fig. 2 contains the results for the inverse susceptibility 
$\chi^{-1}$ against the energy $E$ in the four and zero flavour models. 
Again in these cases, all the points in the Coulomb 
phase are very well fitted by a straight line, which implies $\gamma = 1$ 
independently of the flavour number.

The results of these fits for the critical coupling in the four flavours model 
are $\beta_c= 0.202(3) (8^4)$, $\beta_c= 0.2026(21) (10^4)$. 
The $\chi^{2}
\over{d.o.f.}$ are 2.8, 0.56 respectively, which show the 
high reliability of these fits as well as of the value $\gamma = 1$ for the 
susceptibility critical exponent.

The previous values of $\beta_c$ in the $8^4$ and $10^4$ lattices are on the 
other hand in perfect agreement with the corresponding values 
extracted, following the 
method reported in \cite{NOS}, from the singular behavior of the fermionic 
effective action (see also Table I). As anticipated in the 
introduction, these values are out of the value $0.186(1)$ 
reported in \cite{TED} by eight standard deviations.

The values of the critical couplings 
and energies 
for different lattice sizes and flavour number are reported in Table I, 
where we include also the critical values obtained from the singular 
behavior of the fermion effective action.
As can be seen in this Table, the results for the critical couplings 
extracted from the susceptibility and fermion effective action in the 
two flavour model are also in very good agreement, but they are incompatible 
in the quenched case. Notice however that the quenched value of $\beta_c$ 
obtained from the susceptibility is the same as the one obtained in an 
independent calculation of the chiral order parameter in $24^4$ 
lattices \cite{MARIA} whereas the $\beta_c$ value obtained from the anomalous 
behavior of the fermionic action agrees very well with the critical 
coupling for the monopole percolation transition in the quenched model 
\cite{PERCOLAQ}. 

It is not clear to us the physical origin of the discrepancy 
between the critical couplings obtained by measuring different operators 
in the zero flavour model. However the agreement between critical 
couplings for monopole percolation and singular behavior of the 
fermionic action is, we believe, a clear signal of strong correlation 
between these two phenomena.

We have also done measurements of the longitudinal susceptibility 
$\chi_L$ for massive fermions as well as of the non linear susceptibility 
$\chi_{nl}$, which is related to the four point function and therefore 
to the low energy renormalized coupling. A detailed 
analysis of these 
calculations will be reported in a longer publication. However we report 
in Fig. 3 some results obtained for $\chi^{-1}_L$ as a function of the 
gauge coupling $\beta$ at two values of the fermion mass $m=0.00625, 0.01$. 
The value of the dynamical fermion mass in this figure (the mass in the 
integration measure) has been taken equal 
to zero. 
The inverse chiral susceptibility shows a minimum 
pointing to the critical $\beta$ previously obtained from 
the fit of the results for the massless susceptibility in the Coulomb phase.

Concerning the non linear susceptibility $\chi_{nl}$, we would 
like to anticipate that the results for $\chi_{nl}$ suffer, as 
expected, from strong finite size effects \cite{GNJL}. One of the 
contributions to the massless non linear susceptibility operator can 
be defined as 

$$O_{nl}={2\over V} \sum_i  {1\over{\lambda^4_i}}
\eqno(10)$$

\noindent
where again in this case the sum runs over all positive eigenvalues of 
the massless fermion matrix.

The expression given by equation (10) 
diverges in the thermodynamical limit due to the presence of zero modes 
both in the Coulomb and broken phases. However the divergency in the broken 
phase is much faster because chiral symmetry is spontaneously broken in this 
phase. Then a rough estimate of the critical coupling can be obtained from 
the results for the $v.e.v.$ of the inverse of this operator, 
by defining it as the point where the 
slope of this $v.e.v.$ as a function of the gauge coupling $\beta$ (see Fig. 4)
takes its maximum value. This criterion gives for $\beta_c$ in the quenched 
model a value between $0.25-0.26$ in good agreement with the 
results reported in Table I.

In conclusion we have shown how the analysis of the massless chiral 
susceptibility in the Coulomb phase of noncompact $QED$, even in rather small 
lattices as ours, can be very useful to get precise determinations of 
the critical couplings. 
Indeed what gives high reliability to the susceptibility fits we 
have done in the Coulomb phase is the fact that all the points, except 
those very near to $\beta_c$ which could be affected by finite size effects, 
are very well fitted by a single power law term without higher power 
corrections, even far from $\beta_c$.

The numerical simulations quoted above have been done using the Transputer 
Networks of the Theoretical Group of the Frascati National Laboratories 
and the Reconfigurable Transputer Network (RTN), 
a 64 Transputers array, of the University of Zaragoza.

This work has been partly supported through a CICYT (Spain) - 
INFN (Italy)
collaboration.

\endpage

\endpage
\vskip 1 truecm
\leftline{\bf Figure captions}
\vskip 1 truecm

\noindent
{\bf Fig. 1.} Inverse susceptibility 
in the chiral limit against $\beta$ for the four flavour model, $10^4$ lattice.

\noindent
{\bf Fig. 2.} Inverse susceptibility as a function of the energy $E$ in a 
$10^4$ lattice for zero and four dynamical flavours.

\noindent
{\bf Fig. 3.} Inverse longitudinal susceptibility for $m=0.01$ and $m=0.00625$
against $\beta$ in the quenched (a) and four flavours model (b).

\noindent
{\bf Fig. 4.} $v.e.v.$ of the inverse operator given by eq. (10) against 
$\beta$ in the quenched case, $8^4, 10^4$ and $12^4$ lattices.

\endpage
\vskip 1 truecm
\leftline{\bf Table caption}
\vskip 1 truecm

\noindent
{\bf Table I.} Critical couplings and energies for different lattice sizes 
and flavour numbers. The superscripts $\chi$ and $S$ denote values extracted 
from the chiral susceptibility and effective fermion action respectively.

\endpage

\vbox{\offinterlineskip 
  \hrule
  \halign{&\vrule#&
  \strut\, #\,\cr
  height2pt&\omit&&\omit&&\omit&&\omit&&\omit&&\omit&&\omit&\cr
  &lattice&&\hfil $10^4$\hfil &&\hfil $10^4$\hfil &&\hfil $10^4$\hfil &
   &\hfil $8^4$\hfil &&\hfil $8^4$\hfil &&\hfil $8^4$\hfil &\cr
  height2pt&\omit&&\omit&&\omit&&\omit&&\omit&&\omit&&\omit&\cr
 \noalign{\hrule}
  height2pt&\omit&&\omit&&\omit&&\omit&&\omit&&\omit&&\omit&\cr
  &\# flav. &&\hfil 4\hfil &&\hfil 2\hfil &&\hfil 0\hfil &
   &\hfil 4\hfil &&\hfil 2\hfil &&\hfil 0 \hfil&\cr
  height2pt&\omit&&\omit&&\omit&&\omit&&\omit&&\omit&&\omit&\cr
 \noalign{\hrule}
  height2pt  &\omit&&\omit&&\omit&&\omit&&\omit&&\omit&&\omit&\cr
   &\hfil $E_c^\chi$\hfil &&1.028(8)&&1.027(7)&
    &0.977(12)&&1.034(11)&&1.038(10)&&0.974(8)&\cr
   &\hfil $\beta_c^\chi$ \hfil&&0.2026(21)&&0.2231(18)&
    &0.256(3)&&0.202(3)&&0.221(2)&&0.257(2)&\cr
   &\hfil $E_c^S$\hfil&&1.02(2)&&1.030(15)&&1.030(10)&
    &\hfil&&\hfil&&\hfil&\cr
   &\hfil $\beta_c^S$ \hfil&&0.205(5)&&0.222(3)&
    &0.243(2)&&\hfil&&\hfil&&\hfil&\cr
  height2pt&\omit&&\omit&&\omit&&\omit&&\omit&&\omit&&\omit&\cr}
  \hrule}

\vskip 1.7 truecm
\centerline {Table I}

\end{document}